\documentclass[11pt]{article}
\usepackage{epstopdf}
\usepackage{color}
\usepackage{subfigure}
\usepackage{amsmath}
\usepackage{amssymb}
\usepackage{graphicx,color}
\usepackage{cite}
\usepackage{enumerate}
\usepackage{amsthm}
\usepackage{amsfonts,mathrsfs}
\usepackage{geometry}
\usepackage{ifthen}
\usepackage{graphicx}
\usepackage{float}
\usepackage{bm}
\usepackage[caption=false]{subfig}
\usepackage[utf8]{inputenc}
\usepackage{pifont}

\begin{document}

\title{\bf Coherence dynamics in Simon's quantum algorithm}

\vskip0.1in
\author{\small Linlin Ye$^1$, Zhaoqi Wu$^1$\thanks{Corresponding author. E-mail: wuzhaoqi\_conquer@163.com}, Shao-Ming
Fei$^{2,3}$\\
{\small\it  1. Department of Mathematics, Nanchang University,
Nanchang 330031, China}\\
{\small\it  2. School of Mathematical Sciences, Capital Normal
University, Beijing 100048, China}\\
{\small\it  3.Max Planck Institute for Mathematics in the Sciences,
04103 Leipzig, Germany}}
\date{}
\maketitle

\noindent {\bf Abstract} {\small }\\
Quantum coherence plays a pivotal role in quantum algorithms. We
study the coherence dynamics of the evolved states in Simon's
quantum algorithm based on Tsallis relative $\alpha$ entropy and
$l_{1,p}$ norm. We prove that the coherences of the first register
and the second register both rely on the dimension $N$ of the state
spaces of the $n$ qubit systems, and increase with the increase of
$N$. We show that the oracle operator $O$ does not change the
coherence. Moreover, we study the coherence dynamics in the Simon's
quantum algorithm and prove that in overall the coherence is in
production when $N>4$ and in depletion when $N<4$.

\vskip0.2 in

\noindent {\bf 1 Introduction}\\\hspace*{\fill}\\
Coherence is the crucial physical resource in quantum information
processing and quantum computation \cite{MA}, which plays
significant roles in quantum biology \cite{PMB,LS}, transport theory
\cite{RPM,WBM} and quantum metrology \cite{GVL,GVLQ}. Based on a
rigorous framework \cite{TB}, coherence measures
\cite{XJCM,BKFS,RSP,YCS,ZXN,SLH,MPM,WZZ} and coherence of formation
have been put forward \cite{YXZ,WAY}. The relations between the
coherence and quantum correlations \cite{YYX,XZL,MJY} and phenomena
like freezing coherence \cite{BTR,YXD} have been extensively
studied.

The $l_{1,p}$ norm of coherence, as one of the important coherence
quantifiers, has been proposed in \cite{JYL}, which extends and
unifies the norm-induced coherence measures. Quantifying coherence
based on Tsallis relative $\alpha$ entropy \cite{AS1,AS2} has been
first proposed in \cite{RAEQ}, though it does not satisfy the strong
monotonicity. Zhao and Yu \cite{ZHYC} modified the definition of the
Tsallis relative $\alpha$ entropy based coherence measure and
proposed a well-defined coherence quantifier.

By taking advantage of quantum state superposition and quantum
entanglement, quantum computation can improve the speed of problem
solving\cite{ZhouNR}. The Simon's quantum algorithm provides a clear
exponential gap between the classical and quantum runtime, and plays
a vital role in the evolution of quantum algorithm design
\cite{SDR,CAM}. The first experimental realization of a one-way
implementation of the algorithm has been given in \cite{TMS}. Ghosh
and Sarkar applied the algorithm to the cryptanalysis of several
tweakable enciphering schemes and revealed portions of the secret
keys \cite{GSS}.

The application of coherence in quantum algorithms has attracted
people's attention \cite{PMQ,NMK,HMC,FSH}. Coherence depletion in
quantum algorithm has been studied in \cite{LYC,SHL}. The
complementarity relation between coherence and success probability
of Grover's search algorithm has been investigated in \cite{MPH}.

In this paper, after recalling the Simon's quantum algorithm and
coherence quantifiers, we investigate the coherence dynamics of the
states after each basic operator is applied, and study the coherence
production and depletion. We also give detailed examples to
illustrate the coherence dynamics.

\vskip0.1in

\noindent {\bf 2 The $l_{q,p}$ norm of coherence and Tsallis relative $\alpha$ entropy of coherence in Simon's quantum algorithm}\\\hspace*{\fill}\\
We first review the Simon's quantum algorithm and coherence
quantifiers.

{\it Simon's problem \cite{SDR}} A blackbox function or oracle is
given by a function $f$:\{0,1\}$^{n}\rightarrow$\{0,1\}$^{n}$. It
takes an $n$ bitstring $x=x_{1}$ $x_{2}$ $x_{3}\ldots x_{n}$ as
input, where $x_{i}$ is either zero or one, \{0,1\}$^{n}$ is the
collection of all possible $n$ bitstring. It is assumed that when
$f$ is a many-to-one function, there is a nontrivial $n$-bit string
$s$ such that for any pair of distinct inputs $x$ and $x'$, $f(x)$
and $f(x')$ are equal if and only if $x'=x\oplus s$, where $\oplus$
denotes bitwise modulo 2 addition of two $n$ bit-strings. String
$s=0^{n}$, if $f$ is one-to-one. The goal is to find the secret
string $s$ with the least number of queries.

{\it Simon's quantum algorithm} This is an algorithm to solve the Simon's problem with $O$$(n)$ repetitons of the routine \cite{SDR}:\\
(i) Apply $H^{\otimes n}\otimes I^{\otimes n}$ to the initial state
$|0\rangle^{n}\otimes|0\rangle^{n}$,
where
$H=\frac{1}{\sqrt{2}}\sum_{x,y\in\{0,1\}}(-1)^{xy}|y\rangle\langle
x|$. The resulting state is
\begin{equation}\label{eq1}
|\psi_{H}\rangle=\frac{1}{\sqrt{2^{n}}}\sum_{x\in\{0,1\}^{n}}|x\rangle|0\rangle^{\otimes
n}.
\end{equation}
(ii) Apply the oracle operator $O$ $=$ $\sum_{x,y\in\{0,1\}^{n}}$
$|x\rangle\langle x|\otimes|y\oplus f(x)\rangle\langle y|$ to
compute $f(x)$ and concatenate the answer to $x$. The state (1)
becomes
\begin{equation}\label{eq1}
|\psi_{HO}\rangle=\frac{1}{\sqrt{2^{n}}}\sum_{x\in\{0,1\}^{n}}|x\rangle|f(x)\rangle.
\end{equation}
The value $f(x)$ corresponds to the superposition state
$\frac{|x\rangle+|x\oplus s\rangle}{\sqrt{2}}$ of the first
register. Eq. (2) can be transformed into $\frac{1}{\sqrt{2^{n-1}}}$
$\sum_{x\in R}$ $\frac{|x\rangle+|x\oplus s\rangle}{\sqrt{2}}$
$|f(x)\rangle$, where $R$ represents the coset of $n$ bitstring that satisfies $f(x)=f(x\oplus s)$ in $\{0,1\}^{n}$ with $|R|=2^{n-1}$.\\
(iii) Apply $H^{\otimes n}\otimes I^{\otimes n}$. One gets
$$
\frac{1}{2^{n}}\sum_{x\in R}\sum_{y\in\{0,1\}^{n}}(-1)^{x\cdot
y}\left[1+(-1)^{s\cdot y}\right]|y\rangle|f(x)\rangle.
$$
Note that $1+(-1)^{s\cdot y}=2$ if $s\cdot y\equiv0$ mod 2,
otherwise $1+(-1)^{s\cdot y}=0$. The state can be rewritten as
\begin{equation}\label{eq4}
|\psi_{HOH}\rangle=\frac{1}{2^{n-1}}\sum_{x\in
R}\sum_{y\in\{0,1\}^{n}}(-1)^{x\cdot y}|y,f(x)\rangle,
\end{equation}
where $s\cdot y\equiv0$ mod 2. Denote by $p_{y}$ the probability of
the measurement output state $|y\rangle$. We have
$p_{y}=\left|\frac{1}{2^{n}}(-1)^{x\cdot y}\left[1+(-1)^{s\cdot
y}\right]\right|^{2}
=\frac{1}{2^{n-1}}$, where $x=x_{1}$ $x_{2}$ $x_{3}\ldots x_{n}$ is an input.\\
(iv) Measure the registers. We get the $|(y,f(x))\rangle$ such that $y\cdot s\equiv$ 0 (mod 2).\\

Next, we will discuss the coherence of the states
$|\psi_{H}\rangle$, $|\psi_{HO}\rangle$ and $|\psi_{HOH}\rangle$. To
do this, let us recall some coherence measures. The Tsallis relative
$\alpha$ entropy is defined by \cite{AS1,AS2}
\begin{equation}\label{eq7}
D_{\alpha}(\rho\|\sigma)=\frac{1}{\alpha-1}\left(f_{\alpha}(\rho,\sigma)
-1\right),
\end{equation}
where
$f_{\alpha}(\rho,\sigma)=\mathrm{Tr}(\rho^{\alpha}\sigma^{1-\alpha})$,
$\alpha\in(0,1)\cup(1,\infty)$. With respect to a fixed orthonormal
basis $\{|j\rangle\}_{j=1}^d$ in a $d$ dimensional Hilbert space, a
well-defined coherence quantifier based on Tsallis relative $\alpha$
entropy has been presented when $\alpha\in(0,1)\cup(1,2]$
\cite{ZHYC},
\begin{equation}\label{eq9}
C_{\alpha}(\rho) =\frac{1}{\alpha-1}\left[\sum_{j=1}^{d}\langle
j|\rho^{\alpha}|j\rangle^{\frac{1}{\alpha}}-1\right],
\end{equation}
where $\mathcal{I}$ denotes the set of incoherent states.
$C_{\alpha}(\rho)$ reduces to $\ln2\cdot C_{r}(\rho)$ when
$\alpha\rightarrow1$, where $C_{r}(\rho)=\mathrm{Tr}(\rho\log\rho)-
\mathrm{Tr}(\rho_{\mathrm{diag}}\log\rho_{\mathrm{diag}})$ is the
relative entropy of coherence\cite{TB}, and $C_{\alpha}(\rho)$
reduces to $2\cdot C_{s}(\rho)$ when $\alpha=\frac{1}{2}$, where
$C_{s}(\rho)=1-\sum_{j=1}^{d}\langle j|\sqrt{\rho}|j\rangle^{2}$ is
the skew information of coherence \cite{YCS}.

The $l_{q,p}$ norm of a matrix $A\in M_{n}$ is the $l_{q}$ norm of
the vector formed by the $l_{p}$ norm of the columns of $A$
\cite{JYL},
\begin{equation*}\label{eq10}
l_{q,p}(A)=\left(\sum_{j=1}^{n}l_{p}(A_{j})^{q}\right)^{\frac{1}{q}},~~~1\leq
p,~q\leq\infty,
\end{equation*}
where $A_{j}$ is the columns of $A$, and
$l_{p}(A_{j})=\left(\sum_{i=1}^{n}|A_{i,j}|^{p}\right)^{\frac{1}{p}}$
with $A_{i,j}$ the entry of the $i$th row and $j$th column of $A$.

The coherence based on $l_{q,p}$ norm is a well-defined coherence
measure if and only if $q=1$ and $p\in[1,2]$. The coherence
$C_{1,p}$ for $p\in[1,2]$ is defined by
\begin{equation}\label{eq11}
C_{1,p}(\rho)
=\sum_{j=1}^{n}\left(\sum_{i=1}^{n}|(\rho-\rho_{\mathrm{diag}})_{i,j}|^{p}\right)^{\frac{1}{p}},
\end{equation}
where $i$, $j$ are the $i$th row and $j$th column of the matrix
$\rho-\rho_{\mathrm{diag}}$. Note that when $p=1$, $C_{1,p}(\rho)$
reduces to $ C_{l_{1}}(\rho)$, where $C_{l_{1}}(\rho)=\sum_{i\neq
j}|\rho_{ij}|$ is the $l_{1}$ norm of coherence\cite{TB}.

\vskip0.1in

\noindent {\bf 3 Coherence dynamics in Simon's quantum algorithm}\\\hspace*{\fill}\\
{\bf Theorem 1} The coherence of the state $|\psi_{H}\rangle$ based
on Tsallis relative $\alpha$ entropy and $l_{1,p}$ norm are given by
\begin{equation}\label{eq12}
C_{\alpha}(\rho_{H})=\frac{1}{\alpha-1}\left(N^{1-\frac{1}{\alpha}}-1\right),
\end{equation}
and
\begin{equation}\label{eq12}
C_{1,p}(\rho_{H})=(N-1)^{\frac{1}{p}},
\end{equation}
respectively, where $N=2^{n}$.

${\bf Proof.}$ By Eq. (1), the density operator of
$|\psi_{H}\rangle$ is
\begin{equation}\label{eq13}
\rho_{H}=\frac{1}{2^{n}}\sum_{x,y}|x\rangle\langle y|.
\end{equation}
According to Eqs. (5) and (9), the coherence of $|\psi_{H}\rangle$
based on Tsallis relative $\alpha$ entropy is given by (7).
Combining Eqs. (6) with (9), we obtain the coherence of
$|\psi_{H}\rangle$ based on $l_{1,p}$ norm,
\begin{equation*}\label{eq15}
C_{1,p}(\rho_{H})=N\left(\frac{1}{N^{p}}(N-1)\right)^{\frac{1}{p}}=(N-1)^{\frac{1}{p}}.
\end{equation*}
This completes the proof. $\Box$

{\bf Remark 1} Leting $\alpha=\frac{1}{2}$ and taking the limit
$\alpha\rightarrow1$ in Eq. (7), one has the coherence of $\rho_{H}$
based on skew information and relative entropy,
\begin{equation*}\label{eq16}
C_{s}(\rho_{H})=1-\frac{1}{N}~\mathrm{and}~ C_{r}(\rho_{H})=\log N.
\end{equation*}
Setting $p=1$ in Eq. (8), one gets the $l_{1}$ norm of coherence of
$\rho_{H}$, $C_{l_{1}}(\rho_{H})=N-1$.

The density operator of $|\psi_{HO}\rangle$ is
\begin{equation}\label{eq18}
\rho_{HO}=\frac{1}{2^{n}}\sum_{x,y}|x,f(x)\rangle\langle y,f(y)|.
\end{equation}
According to Eqs. (5), (7) and (10), the coherence of $\rho_{HO}$
based on Tsallis relative $\alpha$ entropy satisfies
\begin{equation}\label{eq20}
C_{\alpha}(\rho_{HO})=C_{\alpha}(\rho_{H}).
\end{equation}
According to Eqs. (6), (8) and (10), the $l_{1,p}$ norm of coherence
of $\rho_{HO}$ satisfies
\begin{equation}\label{eq20}
C_{1,p}(\rho_{HO})=C_{1,p}(\rho_{H}).
\end{equation}
Therefore we obtain the following result.

{\bf Theorem 2} The coherence of the state $|\psi_{HO}\rangle$ based
on the Tsallis relative $\alpha$ entropy and $l_{1,p}$ norm does not
change after applying $O$.

Letting $\alpha=\frac{1}{2}$ and taking the limit
$\alpha\rightarrow1$ in Eq. (11), and setting $p=1$ in Eq. (12), we
have $C_{s}(\rho_{HO})=C_{s}(\rho_{H})$,
$C_{r}(\rho_{HO})=C_{r}(\rho_{H})$ and
$C_{l_{1}}(\rho_{HO})=C_{l_{1}}(\rho_{H})$, respectively.

{\bf Theorem 3} The coherence of the state $|\psi_{HOH}\rangle$
based on Tsallis relative $\alpha$ entropy and $l_{1,p}$ norm are
given by
\begin{equation}\label{eq24}
C_{\alpha}(\rho_{HOH})=\frac{1}{\alpha-1}
\left(4^{\frac{1}{\alpha}-1}N^{2-\frac{2}{\alpha}}-1\right),
\end{equation}
and
\begin{equation}\label{eq25}
C_{1,p}(\rho_{HOH})=\left(\frac{N^{2}}{4}-1\right)^{\frac{1}{p}},
\end{equation}
respectively.

${\bf Proof.}$ The density operator of $|\psi_{HOH}\rangle$ is
\begin{align}\label{eq26}
&\rho_{HOH}=|\psi_{HOH}\rangle\langle\psi_{HOH}|\notag\\
&=\frac{1}{(2^{n-1})^{2}}\sum_{x,x'\in
R}\sum_{y,y'\in\{0,1\}^{n}}|y,f(x)\rangle\langle y'f(x')|.
\end{align}
According to Eqs. (5) and (15), one gets the coherence of
$|\psi_{HOH}\rangle$ based on Tsallis relative $\alpha$ entropy
(13). Combining Eqs. (6) with (15), we obtain the coherence of
$|\psi_{HOH}\rangle$ based on $l_{1,p}$ norm,
\begin{align*}\label{eq27}
C_{1,p}(\rho_{HOH})\notag
=&\frac{N^{2}}{4}\left(\left(\frac{4}{N^{2}}\right)^{p}
\left(\frac{N^{2}}{4}-1\right)\right)^{\frac{1}{p}}\\
=&\left(\frac{N^{2}}{4}-1\right)^{\frac{1}{p}}.
\end{align*}
This completes the proof. $\Box$

{\bf Remark 2} When $\alpha=\frac{1}{2}$ and $\alpha\rightarrow1$ in
Eq. (13), the coherence of $\rho_{HOH}$ based on skew information
and relative entropy are $C_{s}(\rho_{HOH})=1-\frac{4}{N^{2}}$ and
$C_{r}(\rho_{HOH})=\log\frac{N^{2}}{4}$, respectively. Setting $p=1$
in Eq. (14), the $l_{1}$ norm of coherence of $\rho_{HOH}$ is
\begin{equation*}\label{eq29}
C_{l_{1}}(\rho_{HOH})=\frac{N^{2}}{2}-1.
\end{equation*}
Based on Theorem 1 and Theorem 3, the coherence of $\rho_{H}$ and
$\rho_{HOH}$ both rely on the dimensionality $N$, and increase with
$N$.

If the second register is measured first, we will get
$|f(x)\rangle$. Then the state after the measurement is given by
\begin{equation}\label{eq30}
|\psi_{M}\rangle=\frac{1}{\sqrt{2^{n-1}}}\sum_{y\in\{0,1\}^{n}}(-1)^{x\cdot
y}|y\rangle.
\end{equation}
In the following we define the variation of the coherence under an
application in the Simon's quantum algorithm to be
\begin{equation}\label{eq37}
\Delta C(\rho)\equiv C(\rho_{HOH})-C(\rho_{H}).
\end{equation}
Therefore, the coherence is producing when $\Delta$$C(\rho)$$>0$,
and is depleting when $\Delta C(\rho)<0$. Substituting Eqs. (7) and
(13) into (17), and Eqs. (8) and (14) into (17), we obtain the
following results.

{\bf Theorem 4} The variations of coherence based on Tsallis
relative $\alpha$ entropy and $l_{1,p}$ norm during the processing
of the Simon's quantum algorithm are given by
\begin{equation}\label{eq38}
\Delta C_{\alpha}(\rho)=\frac{1}{\alpha-1}\left[\frac{N^{2}}{4}
\left(\frac{4}{N^{2}}\right)^{\frac{1}{\alpha}}-
N\left(\frac{1}{N}\right)^{\frac{1}{\alpha}}\right]
\end{equation}
and
\begin{equation}\label{eq39}
\Delta
C_{1,p}(\rho)=\left(\frac{N^{2}}{4}-1\right)^{\frac{1}{p}}-\left(N-1\right)^{\frac{1}{p}},
\end{equation}
respectively.

{\bf Remark 3} When $\alpha=\frac{1}{2}$ in Eq. (18) and $p=1$ in
Eq. (19), the variations of coherence based on skew information and
$l_{1}$ norm are $\Delta C_{s}(\rho)=\frac{1}{N}-\frac{4}{N^{2}}$
and $\Delta C_{l_{1}}(\rho)=N\left(\frac{N}{4}-1\right)$,
respectively. Accordingly, the variation of relative entropy of
coherence is given by $\Delta C_{r}(\rho)=\log\frac{N}{4}$. It is
shown that the variations of coherence rely on the dimensionality
$N$, and the overall effect is that coherence is in production when
$N>4$ and in depletion when $N<4$ in the Simon's quantum algorithm.

Let us consider some examples to illustrate the coherence dynamics
and the variations of coherence.

{\bf Example 1} Consider dual-bit systems $N=4$. The relations of
bit strings are specified by
$f(|00\rangle)=|00\rangle$, $f(|01\rangle)=|11\rangle$, $f(|10\rangle)=|11\rangle$ and $f(|11\rangle)=|00\rangle$.\\
$\bullet$ (i) Applying $H^{\otimes2}\otimes I^{\otimes2}$ to the
initial state $|00\rangle\otimes|00\rangle$, we get
\begin{equation*}\label{eq44}
|\psi_{H_{1}}\rangle=\frac{1}{2}\left(|00\rangle+|01\rangle+|10\rangle+|11\rangle\right)|00\rangle.
\end{equation*}
Denote $\rho_{H_{1}}=|\psi_{H_{1}}\rangle\langle\psi_{H_{1}}|$. The
coherence of $\rho_{H_{1}}$ based on Tsallis relative $\alpha$
entropy and $l_{1,p}$ norm are
\begin{equation*}\label{eq}
C_{\alpha}(\rho_{H_{1}})=\frac{1}{\alpha-1}\left(4^{1-\frac{1}{\alpha}}-1\right)
\end{equation*}
and
\begin{equation*}\label{eq}
C_{1,p}(\rho_{H_{1}})=3^{\frac{1}{p}},
\end{equation*}
respectively. Accordingly, we have
\begin{equation*}\label{eq}
C_{s}(\rho_{H_{1}})=\frac{3}{4},~~C_{r}(\rho_{H_{1}})=2
~~\mathrm{and}~~C_{l_{1}}(\rho_{H_{1}})=3,
\end{equation*}
respectively.\\
$\bullet$ (ii) Applying the oracle operator $O$, we have
\begin{align}\label{eq46}
|\psi_{HO_{1}}\rangle\notag
=&\frac{1}{2}\left(|00\rangle+|11\rangle\right)|00\rangle\\
+&\frac{1}{2}\left(|01\rangle+|10\rangle\right)|11\rangle.
\end{align}
By simple calculations, the coherence of $|\psi_{HO_{1}}\rangle$ is equal to the coherence of $|\psi_{H_{1}}\rangle$.\\
$\bullet$ (iii) Applying $H^{\otimes2}\otimes I^{\otimes2}$, we
obtain
\begin{align}\label{eq48}
|\psi_{HOH_{1}}\rangle\notag
=&\frac{1}{2}\left(|00\rangle+|11\rangle\right)|00\rangle\\
+&\frac{1}{2}\left(|00\rangle-|11\rangle\right)|11\rangle.
\end{align}
By calculations, we observe that the coherence of
$|\psi_{HOH_{1}}\rangle$ is equal to the coherence of
$|\psi_{HO_{1}}\rangle$. The variation of coherence does not change.

{\bf Example 2} Consider three-qubit systems $N=8$. The three bit
strings are specified by $f(|000\rangle)=|101\rangle$,
$f(|011\rangle)=|110\rangle$, $f(|001\rangle)=|010\rangle$,
$f(|010\rangle)=|000\rangle$, $f(|110\rangle)=|101\rangle$,
$f(|101\rangle)=|110\rangle$,
$f(|111\rangle)=|010\rangle$ and $f(|100\rangle)~~=~~|000\rangle$.\\
$\bullet$ (i) Applying $H^{\otimes3}\otimes I^{\otimes3}$, we have
\begin{align}\label{eq52}
|\psi_{H_{2}}\rangle\notag
=&\frac{1}{2\sqrt{2}}\left(|000\rangle+|001\rangle+|010\rangle\right)|000\rangle\\
+&\frac{1}{2\sqrt{2}}\left(|011\rangle+|100\rangle+|110\rangle\right)|000\rangle\notag\\
+&\frac{1}{2\sqrt{2}}\left(|101\rangle+|111\rangle\right)|000\rangle.
\end{align}
Set $\rho_{H_{2}}=|\psi_{H_{2}}\rangle\langle\psi_{H_{2}}|$. The
coherence based on Tsallis relative $\alpha$ entropy and $l_{1,p}$
norm are
$$
C_{\alpha}(\rho_{H_{2}})=\frac{1}{\alpha-1}\left(8^{1-\frac{1}{\alpha}}-1\right)
$$
and $C_{1,p}(\rho_{H_{2}})=7^{\frac{1}{p}}$, respectively.
Similarly, we obtain $C_{s}(\rho_{H_{2}})=\frac{7}{8}$,
$C_{r}(\rho_{H_{1}})=3$
and $C_{l_{1}}(\rho_{H_{2}})=7$.\\
$\bullet$ (ii) Applying oracle operator $O$, we have
\begin{align}\label{eq53}
|\psi_{HO_{2}}\rangle\notag
=&\frac{1}{2\sqrt{2}}\left(|000\rangle+|110\rangle\right)|101\rangle\\
+&\frac{1}{2\sqrt{2}}\left(|011\rangle+|101\rangle\right)|110\rangle\notag\\
+&\frac{1}{2\sqrt{2}}\left(|001\rangle+|111\rangle\right)|010\rangle\notag\\
+&\frac{1}{2\sqrt{2}}\left(|010\rangle+|100\rangle\right)|000\rangle.
\end{align}
By simple calculations, we have that the coherence of $|\psi_{HO_{2}}\rangle$ is equal to the coherence of $|\psi_{H_{2}}\rangle$.\\
$\bullet$ (iii) Applying $H^{\otimes3}\otimes I^{\otimes3}$, we
deduce that
\begin{align*}\label{eq54}
&|\psi_{HOH_{2}}\rangle\notag\\
&=\frac{1}{4}\left(|000\rangle+|001\rangle+|110\rangle+|111\rangle\right)|101\rangle\notag\\
&+\frac{1}{4}\left(|000\rangle+|111\rangle-|001\rangle-|110\rangle\right)|110\rangle\notag\\
&+\frac{1}{4}\left(|000\rangle+|110\rangle-|001\rangle-|111\rangle\right)|010\rangle\notag\\
&+\frac{1}{4}\left(|000\rangle+|001\rangle-|110\rangle-|111\rangle\right)|000\rangle.
\end{align*}
Let $\rho_{HOH_{2}}=|\psi_{HOH_{2}}\rangle\langle\psi_{HOH_{2}}|$.
The coherence of $\rho_{HOH_{2}}$ based on Tsallis relative $\alpha$
entropy and $l_{1,p}$ norm are
\begin{equation*}\label{eq}
C_{\alpha}(\rho_{HOH_{2}})=\frac{1}{\alpha-1}\left(16^{1-\frac{1}{\alpha}}-1\right)
\end{equation*}
and $C_{1,p}(\rho_{HOH_{2}})=15^{\frac{1}{p}}$, respectively.
Accordingly, we obtain $C_{s}(\rho_{HOH_{2}})=\frac{15}{16}$,
$C_{r}(\rho_{HOH_{2}})=4$ and
$C_{l_{1}}(\rho_{HOH_{2}})=15$.\\
$\bullet$ (iv) The variations of coherence based on Tsallis relative
$\alpha$ entropy and $l_{1,p}$ norm are
\begin{equation*}\label{eq}
\Delta
C_{\alpha}(\rho)=\frac{1}{\alpha-1}\left(16^{1-\frac{1}{\alpha}}-8^{1-\frac{1}{\alpha}}\right)
\end{equation*}
and $\Delta C_{1,p}(\rho)=15^{\frac{1}{p}}-7^{\frac{1}{p}}$,
respectively. Accordingly, we have $\Delta
C_{s}(\rho)=\frac{1}{16}$, $\Delta C_{r}(\rho)=1$ and $\Delta
C_{l_{1}}(\rho)=8$. It is seen that $\Delta C>0$ and the coherence
is producing in the application of the Simon's quantum algorithm.

\vskip0.1in

\noindent {\bf 6 Conclusions}\\\hspace*{\fill}\\
We have studied the coherence dynamics of the evolved states in the
Simon's quantum algorithm. We have shown that the coherences of the
states in the first register and the second both depend on the
dimensionality $N$. They increase with the increase of $N$. We have
proved that the oracle operator $O$ does not change the coherence.
Moreover, we have shown that the variations of coherence rely on the
dimensionality $N$, and the overall effect is that the coherence is
in production when $N>4$ and in depletion when $N<4$. The results in
this paper may shed new light on the study of coherence dynamics in
other quantum algorithms and provide new insights into quantum
algorithm processing tasks.

\vskip0.1in

\noindent

\subsubsection*{Acknowledgements}
\small {Zhaoqi Wu was supported by National Natural Science
Foundation of China (Grant Nos. 12161056) and Natural Science
Foundation of Jiangxi Province (Grant No. 20232ACB211003). Shao-Ming
Fei was supported by National Natural Science Foundation of China
(Grant Nos. 12075159, 12171044), Beijing Natural Science Foundation
(Grant No. Z190005) and the Academician Innovation Platform of
Hainan Province.}


\subsubsection*{Competing interests}
\small {The authors declare no competing interests.}


\subsubsection*{Data availability}
\small {No new data were created or analysed in this study.}



\begin{thebibliography}{S2}

\bibitem{MA}
Nielsen M A and Chuang I L
2000 \textit{Quantum Computation and Quantum Information} (Cambridge: Cambridge University Press)

\bibitem{PMB}
Plenio M B and Huelga S F
2008 \textit{New J. Phys.} \textbf{10} 113019

\bibitem{LS}
Lloyd S
2011 \textit{J. Phys. Conf. Ser.} \textbf{302} 012037

\bibitem{RPM}
Rebentrost P, Mohseni M and Aspuru-Guzik A
2009 \textit{J. Phys. Chem. B} \textbf{113} 9942

\bibitem{WBM}
Witt B and Mintert F
2013 \textit{New J. Phys.} \textbf{15} 093020

\bibitem{GVL}
Giovannetti V, Lloyd S and Maccone L
2011 \textit{Nat. Photonics} \textbf{5} 222

\bibitem{GVLQ}
Giovannetti V, Lloyd S and Maccone L
2004 \textit{Science} \textbf{306} 1330

\bibitem{TB}
Baumgratz T, Cramer M and Plenio M B
2014 \textit{Phys. Rev. Lett.} \textbf{113} 140401

\bibitem{XJCM}
Xu J
2020 \textit{Chin. Phys. B} \textbf{29} 010301

\bibitem{BKFS}
Bu K, Singh U, Fei S-M, Pati A K and Wu J
2017 \textit{Phys. Rev. Lett.} \textbf{119} 150405

\bibitem{RSP}
Rana S, Parashar P and Lewenstein M
2016 \textit{Phys. Rev. A} \textbf{93} 012110

\bibitem{YCS}
Yu C
2017 \textit{Phys. Rev. A} \textbf{95} 042337

\bibitem{ZXN}
Zhu X, Jin Z and Fei S-M
2019 \textit{Quantum Inf. Process.} \textbf{18} 179

\bibitem{SLH}
Shao L, Li Y, Luo Y and Xi Z
2017 \textit{Commun. Theor. Phys.} \textbf{67} 631

\bibitem{MPM}
Piani M, Cianciaruso M, Bromley T R, Napoli C, Johnston N and Adesso G
2016 \textit{Phys. Rev. A} \textbf{93} 042107

\bibitem{WZZ}
Wu Z, Zhang L, Fei S-M and Li-Jost X
2020 \textit{Quantum Inf. Process.} \textbf{19} 125

\bibitem{YXZ}
Yuan X, Zhou H, Cao Z and Ma X
2015 \textit{Phys. Rev. A} \textbf{92} 022124

\bibitem{WAY}
Winter A and Yang D
2016 \textit{Phys. Rev. Lett.} \textbf{116} 120404

\bibitem{YYX}
Yao Y, Xiao X, Ge L and Sun C
2015 \textit{Phys. Rev. A} \textbf{92} 022112

\bibitem{XZL}
Xi Z, Li Y and Fan H
2015 \textit{Sci. Rep.} \textbf{5} 10922

\bibitem{MJY}
Ma J, Yadin B, Girolami D, Vedral V and Gu M
2016 \textit{Phys. Rev. Lett.} \textbf{116} 160407

\bibitem{BTR}
Bromley T R, Cianciaruso M and Adesso G
2015 \textit{Phys. Rev. Lett.} \textbf{114} 210401

\bibitem{YXD}
Yu X, Zhang D, Liu C and Tong D
2016 \textit{Phys. Rev. A} \textbf{96} 060303

\bibitem{JYL}
Jing Y, Li C, Poon E and Zhang C
2021 \textit{J. Math. Phys.} \textbf{62} 042202

\bibitem{AS1}
Abe S
2003 \textit{Phys. Rev. A} \textbf{68} 032302

\bibitem{AS2}
Abe S
2003 \textit{Phys. Rev. A} \textbf{138} 336

\bibitem{RAEQ}
Rastegin A E
2016 \textit{Phys. Rev. A} \textbf{93} 032136

\bibitem{ZHYC}
Zhao H and Yu C
2018 \textit{Sci. Rep.} \textbf{8} 299

\bibitem{ZhouNR}
Zhou N-R, Zhang T-F, Xie X-W and Wu J-Y
2023 \textit{Signal Process.-Image} \textbf{110} 116891

\bibitem{SDR}
Simon D R
1997 \textit{SIAM J. Comput.} \textbf{26} 1474

\bibitem{CAM}
Childs A M and Van Dam W
2010 \textit{Rev. Mod. Phys.} \textbf{82} 1

\bibitem{TMS}
Tame M S, Bell B A, Di Franco C, Wadsworth W J and Rarity J G
2014 \textit{Phys. Rev. Lett.} \textbf{113} 200501

\bibitem{GSS}
Ghosh S and Sarkar P
2021 \textit{Des. Codes Cryptogr.} \textbf{89} 1907

\bibitem{PMQ}
Pan M and Qiu D
2019 \textit{Phys. Rev. A} \textbf{100} 012349

\bibitem{NMK}
Naseri M, Kondra T V, Goswami S, Fellous-Asiani M and Streltsov A
2022 \textit{Phys. Rev. A} \textbf{106} 062429

\bibitem{HMC}
Hillery M
2016 \textit{Phys. Rev. A} \textbf{93} 012111

\bibitem{FSH}
Fu S, He J, Li X and Luo S
2023 \textit{Phys. Scr.} \textbf{98} 045114

\bibitem{LYC}
Liu Y, Shang J and Zhang X
2019 \textit{Entropy} \textbf{21} 260

\bibitem{SHL}
Shi H L, Liu S Y, Wang X H, Yang W L, Yang Z Y and Fan H
2017 \textit{Phys. Rev. A} \textbf{95} 032307

\bibitem{MPH}
Pan M, Situ H and Zheng S
2022 \textit{Europhys. Lett.} \textbf{138} 48002

\end{thebibliography}
\end{document}